\documentclass[galaxies,communication,accept,moreauthors,12pt,a4paper]{mdpi}
\lastpage{x}
\doinum{10.3390/------}
\pubvolume{xx}
\pubyear{2015}
\history{Received: xx / Accepted: xx / Published: xx}

\usepackage{graphicx}
\usepackage[varg]{txfonts}
\usepackage{mathrsfs}
\usepackage{amssymb}

\Title{Isolated Galaxies versus Interacting Pairs with MaNGA}

\Author{Mar\'ia del Carmen Argudo Fern\'andez $^{1,}$*, Fangting Yuan $^{1,\dagger}$, Shiyin Shen $^{1,\dagger}$, Jun Yin $^{1,\dagger}$ Ruixiang Chang $^{1,\dagger}$, and Shuai Feng $^{1,\dagger}$}

\address{%
$^{1}$ Shanghai Astronomical Observatory, Chinese Academy of Sciences, 80 Nandan Road 200030, Shanghai, China}

\contributed{$^\dagger$ These authors contributed equally to this work.}

\corres{margudo@shao.ac.cn.}

\abstract{We present preliminary results of the spectral analysis on the radial distributions of the star formation history in both, a galaxy merger and a spiral isolated galaxy observed with MaNGA. We find that the central part of the isolated galaxy is composed by older stellar population ($\sim$2\,Gyr) than in the outskirts ($\sim$7\,Gyr). Also, the time-scale is gradually larger from 1\,Gyr in the inner part to 3\,Gyr in the outer regions of the galaxy. In the case of the merger, the stellar population in the central region is older than in the tails, presenting a longer time-scale in comparison to central part in the isolated galaxy. Our results are in agreement with a scenario where spiral galaxies are built from inside-out. In the case of the merger, we find evidence that interactions enhance star formation in the central part of the galaxy.}


\begin{document}


\section{Introduction}  \label{Sect:Intro}

From models, it is predicted that disks grow from inner to outer parts of galaxies, the so--called inside--out model of galaxy formation. This means that stars first form in the centre of galaxies and later in their outer regions. As a consequence, we expect to observe gradients in stellar age and metallicity as a function of galactic radius for spiral galaxies \citep{1999ApJ...520...59K,2012A&A...540A..56P}. This effect is difficult to observe from photometry alone because of the age-metallicity degeneracy and the contamination of dust. Therefore, high spatial and spectral resolutions are required. In this regard, \citet{2014A&A...563A..49S} find strong evidences of inside--out growth of the galaxy disk from the study of 7000 ionized regions, extracted from 306 galaxies observed by the CALIFA survey \citep{2012A&A...538A...8S}.

Integral Field Spectroscopy (IFS) is a powerful resource for understanding the physical mechanisms driven the formation and evolution of galaxies, and in particular, processes that drive the star formation. With the advent of the 3D spectroscopy, from IFS data, new observational results are available to explore galaxy assembly models \citep{2013ApJ...764L...1P,2014A&A...570A...6S,2015ApJ...804L..42P}. Integral field unit (IFU) observations provide us with an added dimension to the information available for each galaxy. Clues to the nature of the physical processes that drive star formation in galaxies are encoded in the map. In this regard, MaNGA \citep[Mapping Nearby Galaxies at Apache Point Observatory\footnote{\texttt{http://www.sdss.org/surveys/manga/}};][]{2015ApJ...798....7B} is an integral field spectroscopic survey designed to investigate the internal kinematic structure and composition of gas and stars in an unprecedented sample of nearly 10,000 nearby galaxies. 

First results from the MaNGA prototype run are extremely promising. \citet{2015ApJ...804..125L} have detected radial gradients in the recent star formation histories of galaxies. Their results are consistent with the inside--out picture, but in order to interpret them in light of the predictions from theoretical models, it is necessary to separate the effects of the environment where galaxies are located. Galaxies suffer intrinsic evolution but they are also exposed to the influences of their local and large--scale environments. Do galaxies evolve by nature or by nurture? The observed properties of isolated galaxies are likely to be determined mainly by their initial formation conditions and secular evolutionary processes.

How exactly do galaxies build their disk when they grow in isolation, unaffected by external influences? Isolated galaxies and the power of MaNGA IFU data are key ingredients to reach this goal. We aim at gaining a unique insight into how galaxies formed and evolve by comparing their observed properties with predictions from state--of--the--art chemical evolution models (CEM). Using MaNGA IFUs, we measure their SFH to obtain a much better understanding of disk build--up at local scales. In the talk given in the context of the
European Week of Astronomy and Space Science (EWASS) in Tenerife, June 2015, we presented first analysis and preliminary results of our study. As a test case, under the assumption of the inside-out scenario of galaxy formation, we use the same methodology as for isolated galaxies to explore the SFH in a complete opposite case, merging galaxies. 


\section{Experimental Section}  \label{Sect:Data}

\subsection{Data}

\begin{figure}
\centering{\includegraphics[width=.2\columnwidth]{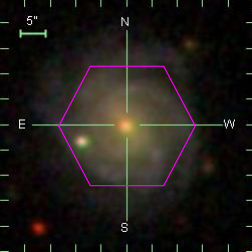}}
\centering{\includegraphics[width=.3\columnwidth]{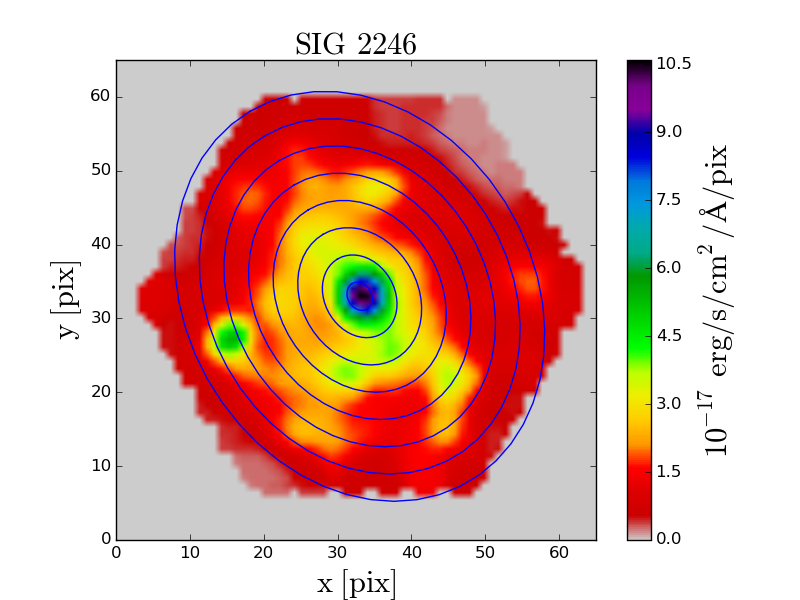}}
\centering{\includegraphics[width=.45\columnwidth]{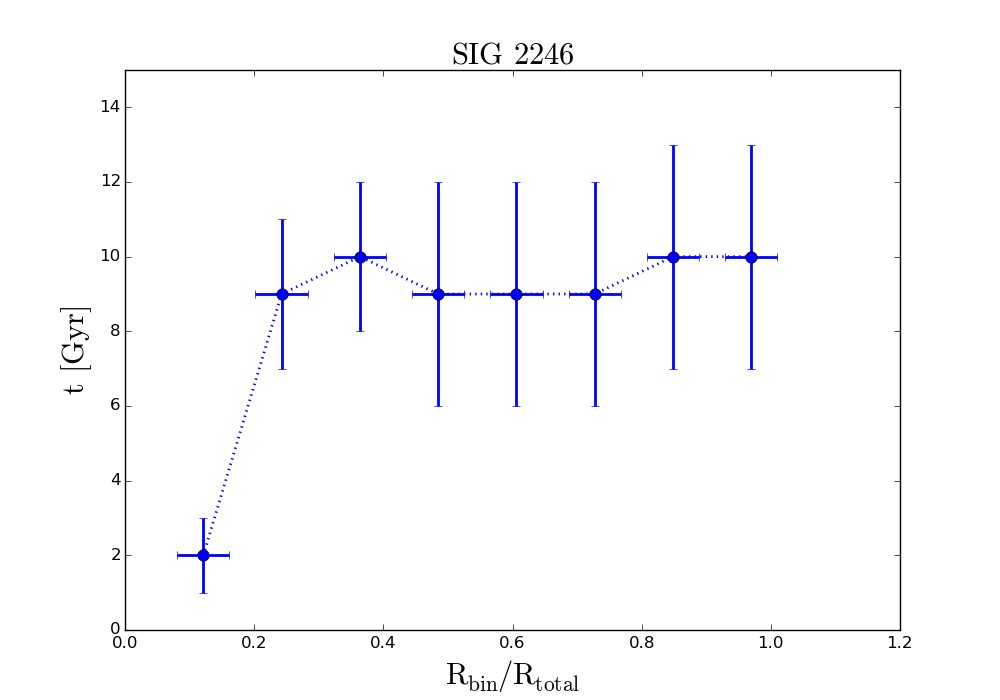}}
\caption{Isolated galaxy SIG 2246. {\it (Left panel):} SDSS three-colour image of the galaxy and the corresponding MaNGA IFU field of view. {\it (Middle panel):} MaNGA H$_{\alpha}$ map of the galaxy and the considered elliptical annuli bins. Flux units are per spatial pixel. {\it (Right panel):} SFH for each radial bin, normalised to the IFU radii. Error bar in the Y-axis corresponds to the $\sigma$ of the Gaussian distribution in each radial bin. Error bar in the X-axis corresponds to radial bin length, manually divided by 3 to separately visualise each error bar (not as a continuum).}
\label{fig:isol}
\end{figure}

The sample of isolated galaxies used in this study is selected from the SDSS-based catalogue of Isolated Galaxies \citep[SIG;][]{2015A&A...578A.110A}, which provided a representative sample of SDSS-selected isolated galaxies for testing galaxy evolution and secular processes in low density regions of the local Universe. These galaxies are isolated with no physical neighbours within a volume of $\Delta\,\varv\,\leq\,500$\,km\,s$^{-1}$ and $d~\leq~1$\,Mpc projected radius, where $\Delta\,\varv$ is the absolute value of the line-of-sight velocity difference between the galaxy and its neighbours. In particular, we show preliminary results for SIG 2246, an almost face-on spiral isolated galaxy that has been already observed with MaNGA (see left panel in Fig.~\ref{fig:isol}). To study the radial distribution of the SFH we consider the stacked spectra in eight radial bins within 2\,arcsec major axis elliptical annuli, as it is shown in the middle panel of Fig.~\ref{fig:isol}, over-plotted in the H$_{\alpha}$ map of the galaxy, which traces star formation. We consider ellipticity of the galaxy according to its value in the NASA-Sloan Atlas (NSA; Blanton M. \href{url}{http://www.nsatlas.org}) catalogue.

\begin{figure}
\centering{\includegraphics[width=.2\columnwidth]{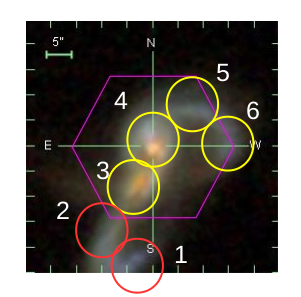}}
\centering{\includegraphics[width=.3\columnwidth]{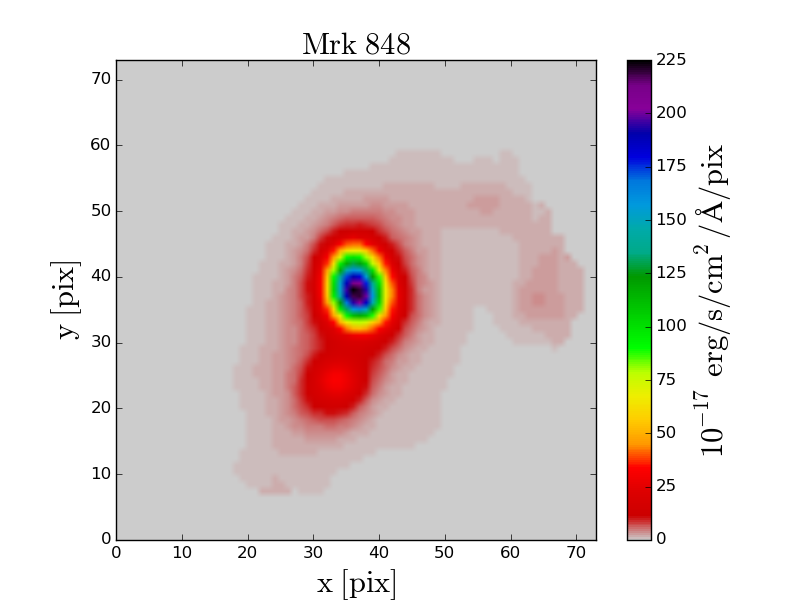}}
\centering{\includegraphics[width=.45\columnwidth]{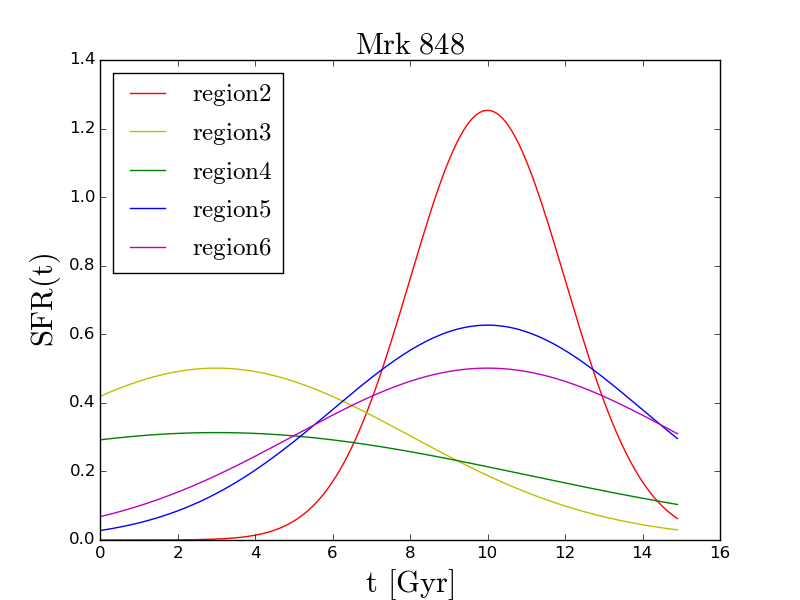}}
\caption{Interacting pair Mrk 848. {\it (Left panel):} SDSS three-colour image of the object. Yellow circles correspond to the regions completely covered by the MaNGA IFU, meanwhile regions with red circles correspond to regions not covered or partially covered. {\it (Middle panel):} MaNGA H$_{\alpha}$ map of the galaxy merger. Flux units are per spatial pixel. {\it (Right panel):} Estimated SFH for each selected region in colour-code according to the legend.}
\label{fig:merger}
\end{figure}

In order to test the methodology when considering galaxies in interaction, we selected Mrk 848, an observed galaxy merger listed in the MaNGA ancillary program on observing close pairs, which will be the largest IFU sample for pair/interacting galaxies in the nearby Universe. This galaxy merger shows two well defined tails (see the left panel in Fig.~\ref{fig:merger}), one in the north and other in the south, where the two nuclei are well identified in the central part. As it is shown in the left panel of Fig.~\ref{fig:merger}, the MaNGA IFU do not cover the full object, we therefore divided the merger into 6 regions. Yellow circles in the left panel of Fig.~\ref{fig:merger} correspond to the regions where the MaNGA IFU completely covers the object, meanwhile red circles correspond to the regions out the IFU. Even if the region 2 is partially covered, we also considered the stacked spectra in that region to explore the SFH in the south tail.

\subsection{Methodology}

To elucidate how galaxies grow with time, one needs to recover the SFH, both in space and time, for individual galaxies. To constraint the SFH for the galaxies considered in our study we apply the so--called ``fossil record method''. Physical properties of the stellar population of the observed spectra (i. e., age, metallicity, and dust attenuation) are then determined by comparing to a modelled spectra. To create a modelled spectral energy distribution (SED) we apply the evolutionary synthesis modelling technique, considering that the SFH follows a determined distribution. For the purpose of this study, based on the inside--out model of galaxy formation, we assume that the SFR of the galaxy vary with time following a Gaussian distribution \citep{1999A&A...350...38C}.

Since a single stellar population (SSP) does not reproduce well the stellar spectra, we use full spectrum modelling to model the SED as a combination of different SSPs, hereafter composite stellar population (CSP). Several models that provide SSPs can be found in the literature, one of the most commonly used is the model developed by \citet{2003MNRAS.344.1000B}, better known as the BC03 model. We adopted this model, assuming a Salpeter IMF \citep{1955ApJ...121..161S} and we fix to solar metallicity as the starting point for this pilot study. Then, we model 100 CSPs where the SFH is parametrized by two free parameters $t_0$ and $\tau$, from 1 to 10 Gyr, which are the mean and sigma values of the Gaussian respectively. 

The selection of a Gaussian distribution will allow an easier comparison of the physical parameters with the theoretical models. Then $t_0$ represents the mean age of the stellar population and $\tau$ represents the time-scale. We therefore estimate $t_0$ and $\tau$ by applying the fossil record method to the stacked spectra in each radial bin for the isolated galaxy, and to the stacked spectra in each region of the merger.

It is important to note that some sources of uncertainties arise when mapping of observed spectra to astrophysical information. Uncertainties come from the observed spectra (noise and calibration of the data), from synthesis models, and from the limitations of the spectral synthesis method \citep{2014A&A...561A.130C}. Moreover, the information extracted from the multi stellar population modelling differs among implementations \citep{2015arXiv150908552S}. Since we present here a pilot study and some improvements will be implemented in modelling and full spectral fitting, we leave the analysis of the uncertainties for the future work.


\section{Results and Discussion}  \label{Sect:Res}

\subsection{SIG 2246}

Following the methodology mentioned above, we constraint the SFH of the isolated galaxy SIG 2246. As it is shown in the right panel in Fig.~\ref{fig:isol}, we estimate the mean age of the average stellar population (blue points), in each radial bin considered, and the corresponding time-scale (error bars in the Y-axis). 

In average, the central part of the galaxy is composed by older stellar population ($\sim$2\,Gyr) than in the outskirts ($\sim$5\,Gyr younger), where the most recent episode of star formation is not younger than 12\,Gyr. Also, the time-scale is gradually larger from 1\,Gyr in the inner part to 3\,Gyr in the outer regions of the galaxy. This result is in agreement with the inside-out model of galaxy formation introduced in Sect.~\ref{Sect:Intro}.

\subsection{Mrk 848}

We apply the same methodology to study the evolution of SIG 2246, to constraint the SFH in the galaxy merger Mrk 848. The left panel in Fig.~\ref{fig:merger} shows the estimated SFH in each considered region of the object. The yellow and green Gaussian distributions correspond to the SFH in the centre (regions 3 and 4). The resulted SFH for regions 2, 5, and 6 correspond to the red, blue, and magenta distributions.

The stellar population in the central region is older than in the tails. If we base our interpretation on spiral galaxies forming under the inside-out model, then regions 3 and 4 correspond to the nuclei of each original spiral galaxy before merging. Therefore, the stellar populations in regions 2, 5, and 6 correspond, in average, to their spiral arms. Moreover, we find that they have the same mean stellar age, about 10\,Gyr but different time-scales, where it is longer in the north tail. 

We also find a long time-scale in the centre of the merger, even larger than the time-scale in the case of the study of SIG 2246. In comparison with the short time-scale in the isolated galaxy, this result suggests that interactions enhance the star formation in the central part of the galaxies.


\section{Conclusions}

We present preliminary results of the spectral analysis on the radial distributions of the star formation history in both, a galaxy merger and a spiral isolated galaxy observed with MaNGA. Our main conclusions are the following:

\begin{enumerate}
 \item We presented preliminary results of the study of the evolution of SIG 2246, an almost face-on spiral isolated galaxy, using IFS from MaNGA, which is up to now the largest IFU survey for galaxies in the local Universe. We find a radial gradient in the SFH of SIG 2246. Our results are in agreement with a scenario where spiral galaxies are built from inside-out.
 \item We also apply the same methodology to study the SFH in the galaxy merger Mrk 848. We find evidence that interactions keep star formation in the central part of the galaxies.
\end{enumerate}

Galaxies present gradients in SFH but also in their chemical enrichment history. According to this, we should take into account that our preliminary results could be biased by the fact that we modelled our CSPs at a fixed solar metallicity. Different metallicities will be considered in the synthesis modelling for the future work.

We expect, for the very first time, to find some clues on how spiral galaxies form and evolve. To reach this goal, we will combine the observed galaxy properties, as result from the spectral fitting techniques, with the predictions from CEMs. First conclusions are expected after comparing the constrained SFH from models to the observed SFH. Results will elucidate what is the most plausible formation scenario. Therefore, we will not only say if galaxies formed by inside-out or not, we will also quantify it by comparing with CEMs. 


\acknowledgments{Acknowledgments}

MAF acknowledges Zhong Jing for his help on first steps in the development of the spectral fitting code. MAF is grateful for financial support from PIFI (funded by Chinese Academy of Sciences President's International Fellowship Initiative) Grant No. 2015PM056. YFT is a LAMOST fellow and this work is supported by NSFC with No. 11303070 (PI : YFT), No. 11433003 (PI Shu Chenggang), and No.11173044 (PI HJL). This work was partly supported by the Strategic Priority Research Programme ``The Emergence of Cosmological Structures'' of the Chinese Academy of Sciences (CAS; grant XDB09030200), the National Natural Science Foundation of China (NSFC) with the Project Number of 11433003 and the ``973 Programme'' 2014 CB845705.

Funding for the Sloan Digital Sky Survey IV has been provided by
the Alfred P. Sloan Foundation, the U.S. Department of Energy Office of
Science, and the Participating Institutions. SDSS-IV acknowledges
support and resources from the Center for High-Performance Computing at
the University of Utah. The SDSS web site is www.sdss.org.

SDSS-IV is managed by the Astrophysical Research Consortium for the 
Participating Institutions of the SDSS Collaboration including the 
Brazilian Participation Group, the Carnegie Institution for Science, 
Carnegie Mellon University, the Chilean Participation Group, the French Participation Group, Harvard-Smithsonian Center for Astrophysics, 
Instituto de Astrof\'isica de Canarias, The Johns Hopkins University, 
Kavli Institute for the Physics and Mathematics of the Universe (IPMU) / 
University of Tokyo, Lawrence Berkeley National Laboratory, 
Leibniz Institut f\"ur Astrophysik Potsdam (AIP),  
Max-Planck-Institut f\"ur Astronomie (MPIA Heidelberg), 
Max-Planck-Institut f\"ur Astrophysik (MPA Garching), 
Max-Planck-Institut f\"ur Extraterrestrische Physik (MPE), 
National Astronomical Observatory of China, New Mexico State University, 
New York University, University of Notre Dame, 
Observat\'ario Nacional / MCTI, The Ohio State University, 
Pennsylvania State University, Shanghai Astronomical Observatory, 
United Kingdom Participation Group,
Universidad Nacional Aut\'onoma de M\'exico, University of Arizona, 
University of Colorado Boulder, University of Oxford, University of Portsmouth, 
University of Utah, University of Virginia, University of Washington, University of Wisconsin, 
Vanderbilt University, and Yale University.


\authorcontributions{Author Contributions}

M.A., F.Y., S.S., and J.Y. conceived and designed the experiments; M.A. performed the experiments; M.A., S.S., F.Y., J.Y, and R.C. analysed the data; M.A., F.Y., J.Y, R.C., and S.F. contributed materials/analysis tools; M.A. wrote the paper.


\conflictofinterests{Conflicts of Interest}

The authors declare no conflict of interest.

\bibliographystyle{mdpi}
\makeatletter
\renewcommand\@biblabel[1]{#1. }
\makeatother


%


%

\end{document}